\documentclass[12pt]{article}

\usepackage{sbc-template}
\usepackage{graphicx,url,amsmath}
\usepackage[utf8]{inputenc}

\usepackage{orcidlink}

\hypersetup{
    pdfborder={0 0 0}
}

\title{Evaluating Parameter Transfer in FALQON Across Graph Families}

\author{Alisson dos Passos Fumaco\inst{1}\orcidlink{0009-0003-3258-5146}, Marcos Vinicius Reballo\inst{1}\orcidlink{0009-0003-6396-9302},\\ Fernando Augusto Caletti de Barros\inst{1}\orcidlink{0009-0008-5152-7036}, Gabriel Fernandes Thomaz\inst{1}\orcidlink{0009-0007-7410-5189}, \\  Eduardo I. Duzzioni\inst{1,2}\orcidlink{0000-0002-8971-2033} }

\address{Instituto de Pesquisas Eldorado
  -- Porto Alegre -- RS -- Brazil
\nextinstitute
  Departamento de Física\\
 Universidade Federal de Santa Catarina (UFSC), 88040-900 -- Florianópolis, SC -- Brazil
  \email{\{alisson.passos,marcos.reballo,fernando.barros\}@eldorado.org.br}
  \email{\{gabriel.thomaz,eduardo.duzzioni.BE\}@eldorado.org.br}
  \email{eduardo.duzzioni@ufsc.br}
}

\begin{document} 

\maketitle

\begin{abstract}
We evaluate FALQON parameter transfer for Max-Cut, transferring sequences from small donors ($n \in \{8,10,12\}$) to 14-node recipients. Using 3-regular and Erd\H{o}s--R\'enyi families, we show that transfer success is dictated by the recipient graph, not the donor. Transfer excels for dense recipients—achieving high approximation ratios regardless of the donor—but remains challenging in sparse cross-family cases. Crucially, performance is highly resilient to donor size, with 8-node donors matching larger instances. Thus, cheap small graphs can provide robust parameters for larger targets, significantly reducing the measurement overhead of the feedback loop.
\end{abstract}

\section{Introduction}

Near-term quantum devices have several issues that prevent them from being used to execute algorithms applied to practical problems. In this context, Variational Quantum Algorithms (VQAs) have emerged as a leading paradigm for exploiting the currently available quantum devices, particularly in the context of combinatorial optimization and quantum simulation. Among these, the Quantum Approximate Optimization Algorithm (QAOA) is attracting substantial attention due to its demonstrated performance across a range of optimization problems encoded as Ising Hamiltonians through QUBO formulation \cite{farhi2014qaoa, Lucas2014IsingReview, zhou2020qaoa}. At the same time, ideas from control theory have increasingly been used to design new quantum optimization strategies, emphasizing feedback, stability, and dynamical structure \cite{Brif2010Control, DongPetersen2013LyapunovSurvey, Grivopoulos2003Lyapunov}.

A promising example of this convergence of quantum optimization and control theory is the Feedback-based Algorithm for Quantum Optimization (FALQON), which reformulates variational optimization as a closed-loop control problem \cite{magann2022falqon}. In this approach, instead of relying on classical gradient descent over a fixed-depth ansatz, FALQON iteratively constructs the quantum circuit by measuring a control observable extracted from the cost Hamiltonian and updating a feedback gain that governs the next layer of the circuit \cite{magann2022lyapunov}. In this way, it is possible to completely eliminate the explicit classical optimization loop and provide a mechanism for driving the system toward low-energy states.

Despite the mentioned advantages, FALQON faces a fundamental challenge shared with other depth-adaptive VQAs: current attempts to apply FALQON to real-world problems have resulted in high-depth circuits, suggesting that depth increases with problem size and desired solution quality. This means that, as the number of required qubits grows, the necessary number of measurements makes calculating the control parameters increasingly expensive. This challenge motivates the exploration of new methods and strategies that could improve the algorithm scalability for larger problems by reusing or transferring information across related problem instances to mitigate the optimization overhead by using smaller problems to compute the control parameters.

In FALQON, the analogue of variational angles is the sequence of feedback gains, which determine how the control Hamiltonian is applied at each iteration. These gains are not arbitrary optimization variables. Instead, they arise from measured expectation values and encode dynamical information linking the current quantum state to the cost Hamiltonian \cite{magann2022falqon}. From a control-theoretic perspective, they reflect system-specific response characteristics and therefore may generalize across problem families with shared structural features.

The central motivation of this work is to investigate whether such feedback parameters can be transferred between related FALQON instances -- across increasing problem sizes or within ensembles of structurally similar Hamiltonians -- in order to reduce the optimization time, suppress transient instability, and accelerate convergence. If successful, parameter transfer would significantly enhance the scalability and practicality of FALQON for large combinatorial problems.

% \subsection{Contributions}
% - Systematic investigation of parameter transfer in FALQON across:
%  **Problem sizes** $n, n+\Delta n$.
% **Hamiltonian families**: RRG(k) and Erdos–Renyi $G(n,p)$.
In this work, we present a systematic study of parameter transferability in quantum optimization algorithms based on FALQON. Our main contributions are as follows:

\begin{enumerate}
    \item \textbf{Problem-size transferability:} We analyze the reuse of FALQON control parameter learned on instances of size $n \in \{8,10,12\}$ when initializing or guiding optimization on larger instances with $n = 14$, quantifying performance degradation and convergence behavior.
    \item \textbf{Hamiltonian family dependence:} We investigate transfer across canonical random graph ensembles, including random regular graphs RRG(\emph{k}) and Erdõs-RÉnyi graphs $G(n,p)$, which are widely used benchmarks in QAOA and graph optimization studies \cite{Bollobas2001RandomGraphs, Herrman2021QAOAMaxCutStructures}.
\end{enumerate}

% TODO: MOVE line below to conclusion section
%By situating FALQON within the growing literature on parameter transfer and warm-starting in quantum optimization, this work extends these ideas to feedback-based, control-inspired algorithms and provides new insights into their scalability and generalization properties. In parallel, with the executed experiments, it was possible to show a way to improve FALQON performance by using Parameter Transfer while still achieving the same result quality as the original FALQON implementation.

The information presented in this work is organized in a sequence of 5 sections, starting with \textbf{Section \ref{sec:rw}}, in which a summary of the current state-of-the-art of Variational and Feedback-based algorithms is presented, the landscape of Parameter Transfer experiments on Quantum Optimization, and the known challenges faced when scaling and generalizing Quantum Algorithms for practical problems. Then, in \textbf{Section \ref{sec:preliminaries}}, the base information required to proceed with the discussions brought by the experiments conducted in this work is presented, describing the FALQON formalism and specific characteristics of Max-Cut problems mapped to Hamiltonians, along with the specific definitions of random graph ensembles considered in the experiments. After that, in \textbf{Section \ref{sec:methodology}}, the framework used for the investigation and experiments is described, along with the specific experiments chosen to evaluate the characteristics of Parameter Transfer used with FALQON in different scenarios. In \textbf{Section \ref{sec:rd}}, the achieved results are presented, along with all the specific information related to each execution and all the findings brought by each of them. After presenting the experiments, the results are explored, analyzing the Problem-size Transferability and the Hamiltonian Family dependence on different scenarios. By the end, in \textbf{Section \ref{sec:conclusion}}, the main findings of the work are presented, providing more information about transfer quality across recipient regimes, findings about the transfer effectiveness resilience to the size gap between donor and recipient graphs, and showing that Parameter Transfer is a promising strategy to enhance FALQON-based optimization.

\section{Related Work}
\label{sec:rw}

% Hybrid quantum algorithms have emerged as leading candidates for exploiting near-term quantum devices in combinatorial optimization. Among these approaches, \emph{variational quantum algorithms} (VQAs) combine shallow parameterized circuits with classical optimization routines. Despite their conceptual appeal, VQAs face well-known challenges that hinder scalability and practical robustness. Motivated by these limitations, alternative strategies leveraging ideas from quantum control and measurement-based feedback have been proposed. 

Hybrid quantum algorithms have emerged as leading candidates for exploiting near-term quantum devices in combinatorial optimization. Among these approaches, \emph{variational quantum algorithms} (VQAs) combine shallow parameterized circuits with classical optimization routines. Despite their conceptual appeal, VQAs face well-known challenges that hinder scalability and practical robustness, such as the barren plateau phenomenon and hardware-induced noise \cite{Cerezo2021VQA, mcclean2018barren}. Motivated by these limitations, alternative strategies leveraging ideas from quantum control and measurement-based feedback have been proposed.

Accordingly, this section reviews related work in three main areas: the formulation and properties of variational and feedback-based algorithms (Section~\ref{sec:rw31}); parameter transfer techniques exploiting structural regularities across problem instances (Section~\ref{sec:rw32}); and strategies addressing scalability, generalization, and optimization stability (Section~\ref{sec:rw33}).

\subsection{Variational \& Feedback-Based Algorithms}
\label{sec:rw31}

Variational quantum algorithms constitute a central paradigm for near-term quantum optimization, with the \emph{Quantum Approximate Optimization Algorithm} (QAOA) as a canonical example~\cite{farhi2014qaoa,zhou2020qaoa}. QAOA alternates between cost and mixer Hamiltonians in a parameterized circuit, relying on a classical outer loop to navigate non-convex energy landscapes. However, practical performance is often limited by challenging parameter optimization, the presence of barren plateaus, and sensitivity to noise, particularly at increasing circuit depths.

To mitigate these limitations, algorithms inspired by Lyapunov control introduce deterministic, sequential parameter updates. A prominent example is the \emph{Feedback-Based Algorithm for Quantum Optimization} (FALQON), which assigns circuit parameters iteratively using expectation values of commutators between the problem and driver Hamiltonians~\cite{magann2022falqon}. This construction guarantees a monotonic decrease of the cost Hamiltonian expectation value and eliminates the need for a classical optimizer, shifting the focus to locally informed, dynamics-driven parameter assignment while retaining a circuit structure closely related to QAOA.

\subsection{Parameter Transfer in Quantum Optimization}
\label{sec:rw32}

A growing body of work has shown that optimal parameters of variational quantum algorithms often exhibit significant structure, including smoothness and concentration around characteristic values for classes of related problem instances. In the context of QAOA, this has led to \emph{parameter transfer}, where parameters optimized for a given instance are reused as initializations or fixed schedules for structurally similar problems. Early studies demonstrated successful transfer between graph instances sharing local structural properties, such as degree distribution, with only minor degradation in performance~\cite{galda2023similarity}. 

Parameter reuse across related problem instances has been extensively studied. For instance, near-optimal parameters for weighted Max-Cut have been shown to transfer across graph sizes and weight patterns with minimal performance degradation using suitable ansatzes and rescaling heuristics~\cite{Shaydulin2022ParamTransfer}, and related data-driven approaches exploit previously optimized instances to avoid re-optimization~\cite{Akshay2021VQAInformationSharing,Li2023DataDrivenQAOA}. Subsequent work leveraged graph representation learning to systematically identify suitable donor instances, further improving robustness~\cite{falla2024graph}. 
%PROPOSAL
These approaches leverage the observation that variational parameters implicitly encode structural properties of the cost Hamiltonian \cite{Wang2018FermionicQAOA, Herrman2021QAOAMaxCutStructures}.
%PROPOSAL
Theoretical analyses have complemented these findings by identifying symmetries and transferable domains in the QAOA parameter landscape, clarifying when and why parameter reuse is effective~\cite{lyngfelt2025symmetry}. 

In contrast, existing approaches to parameter reuse in feedback-based optimization rely on learning-mediated strategies. Pérez \emph{et al.} trained classical machine learning models to predict FALQON parameter schedules, reporting accurate performance for instances close to the training distribution but reduced generalization to larger graphs~\cite{perez2026learning}. This motivates the exploration of feedback-inspired optimization schemes that enable direct and scalable parameter transfer without reliance on machine learning models.

\subsection{Scalability and Generalization in Quantum Algorithms}
\label{sec:rw33}

Scalability and generalization pose fundamental challenges for near-term quantum optimization algorithms. Although increasing circuit depth can improve solution quality, deeper variational circuits exacerbate optimization difficulty and sensitivity to noise, often leading to unstable training dynamics.

One successful strategy to mitigate these effects is \emph{warm-starting}, where quantum optimization algorithms are initialized using approximate solutions obtained from classical heuristics or relaxations. In the context of QAOA, warm-starting significantly enhances performance at low depths by biasing the initial state toward promising regions of the solution space~\cite{egger2021warmstart}, with the classical approximation quality strongly correlating with quantum performance~\cite{okada2023systematic}. 

Beyond initialization, several works have examined the stability of variational optimization under realistic noise and finite sampling, showing that noise can significantly distort optimization landscapes and hinder convergence. These observations reinforce the motivation for optimization schemes that reduce reliance on iterative classical optimization and instead exploit structural priors.

Taken together, these works position FALQON as a compelling alternative to fully variational optimization for near-term quantum devices. By replacing global classical optimization with feedback-driven, dynamics-informed parameter updates, FALQON mitigates key limitations of variational quantum algorithms, including optimization instability and noise sensitivity~\cite{magann2022falqon,magann2022lyapunov}, while opening new avenues for robust parameter transfer.

\section{Preliminaries}
\label{sec:preliminaries}
\subsection{FALQON formalism}
\label{subsec:falqon-formalism}

FALQON~\cite{magann2022falqon} is governed by a time-dependent Schrödinger equation, $i \frac{d}{dt} |\psi(t)\rangle = (H_C + \beta(t) H_M) |\psi(t)\rangle$, where the cost Hamiltonian $H_C$ encodes the problem, the mixer Hamiltonian $H_M$  drives transitions, and $\beta(t)$ is a real-valued control signal.

FALQON employs a Lyapunov control strategy to monotonically decrease the expected cost $V(t) = \langle \psi(t) | H_C | \psi(t) \rangle$. Differentiating $V(t)$ yields $\frac{d}{dt} V(t) = A(t) \beta(t)$, where $A(t) = \langle \psi(t) | i[H_M, H_C] | \psi(t) \rangle$. To guarantee $\frac{d}{dt} V(t) \le 0$, FALQON sets the feedback law to $\beta(t) = -A(t)$~\cite{Brif2010Control,DongPetersen2013LyapunovSurvey,magann2022falqon}.

For gate-based quantum hardware, this continuous evolution is discretized into $L$ steps of duration $\Delta t$. The circuit is approximated by alternating unitaries $U_C = \exp(-i H_C \Delta t)$ and $U_M(\beta_k) = \exp(-i \beta_k H_M \Delta t)$, yielding the overall unitary
\begin{equation}
  U(\boldsymbol{\beta}) = \prod_{k=1}^L U_M(\beta_k) U_C.
\end{equation}
At each step $k$, the commutator expectation $A_k = \langle \psi_k | i[H_M, H_C] | \psi_k \rangle$ is measured to determine the next control parameter via the update rule $\beta_{k+1} = -A_k$. This ensures the discrete-time cost decreases monotonically across layers, subject to finite sampling noise~\cite{magann2022falqon}.

\subsection{Max-Cut and random graph ensembles}
\label{subsec:max-cut}

The Max-Cut problem on an undirected, unweighted graph $G = (V,E)$, seeks a vertex bipartition that maximizes the number of crossing edges. By assigning spin variables $z_i \in \{+1,-1\}$ to each vertex, the contribution of an edge $(i,j)$ to the cut is $\frac{1}{2}(1 - z_i z_j)$. Promoting these classical spins to Pauli $Z$ operators yields the cost Hamiltonian~\cite{Lucas2014IsingReview}:
\begin{equation}
  H_C = \frac{1}{2} \sum_{(i,j) \in E} \bigl( \mathrm{1} - Z_i Z_j \bigr).
\end{equation}
The ground state of $H_C$ encodes the maximum cut of $G$, and this operator serves as the cost Hamiltonian in our experiments.

Following standard practice in the literature, we consider two random graph ensembles as target instances~\cite{farhi2014qaoa,Wang2018FermionicQAOA,Herrman2021QAOAMaxCutStructures,magann2022falqon,Bollobas2001RandomGraphs}:
\begin{itemize}
    \item \textbf{3-Regular Graphs ($\mathcal{G}_{n,3}$):} Graphs sampled uniformly where every vertex has degree exactly three, providing highly symmetric test cases.
    \item \textbf{Erdős--Rényi Graphs ($G(n,p)$):} Graphs where each of the $\binom{n}{2}$ possible edges exists independently with probability $p$, allowing us to probe performance across a broad range of densities.
\end{itemize}

\section{Methodology}
\label{sec:methodology}

We investigate parameter transfer in FALQON using a donor--recipient protocol, where a parameter sequence $\boldsymbol{\beta}^{(D)} = (\beta^{(D)}_1,\ldots,\beta^{(D)}_L)$ optimized on a smaller \emph{donor} (D) graph is directly applied to a moderately larger \emph{recipient} (R) graph. We employ a strict one-to-one layer mapping, setting $\beta^{(R)}_t = \beta^{(D)}_t$ for $t=1,\ldots,L$. This position-wise transfer preserves the temporal structure of the learned control sequence without any layer re-indexing, interpolation, or post-transfer adjustments.

To assess how well parameters generalize across sizes and structural connectivities, donor graphs are generated with sizes $n \in \{8,10,12\}$, while recipient graphs are fixed at $n' = 14$. We evaluate transfers within and across two random graph families: 3-regular graphs and Erd\H{o}s--R\'enyi graphs $G(n,p)$. For the Erd\H{o}s--R\'enyi family, donor instances use edge probabilities $p \in \{0.2, 0.3, 0.4, 0.5\}$, whereas recipient instances span $p \in \{0.2, 0.3, \ldots, 1.0\}$. This comprehensive evaluation allows us to isolate the effects of donor size and structural similarity on transfer quality.

For each configuration, the transferred parameters are evaluated on 10 independently sampled recipient graphs, and the results are averaged. We assess the viability of the transfer using the approximation ratio, which measures the solution's cut value relative to the exact Max-Cut optimum. To quantify how well the control behavior is preserved, the mean approximation ratio achieved on the recipient instances is directly compared against the reference performance obtained on the original donor graph.

All quantum algorithmic evaluations are conducted via exact state-vector simulation. On real quantum devices, the sequential nature of FALQON requires a continuous classical feedback loop to estimate the commutator expectation values. This loop faces significant challenges, such as finite sampling noise, readout errors, and communication latency between the quantum processor and classical hardware. Therefore, using exact simulation ensures that the feedback dynamics are evaluated deterministically, isolating the fundamental transferability of the parameter sequences from these hardware-induced overheads and decoherence. Furthermore, the maximum recipient size of $n'=14$ serves two crucial purposes. First, it allows us to exactly compute the global Max-Cut optimum for every generated instance using classical brute-force evaluation, establishing an absolute performance baseline devoid of heuristic approximations.

\section{Results and Discussion}
\label{sec:rd}

We organize the analysis around three empirical patterns observed for 14-node recipient graphs. The discussion focuses on the relationship between the transferred approximation-ratio trajectories and the donor-side training trajectory to characterize whether FALQON parameter sequences preserve, degrade, or improve after transfer.

\begin{figure}[htb]
    \centering
    \includegraphics[width=\columnwidth]{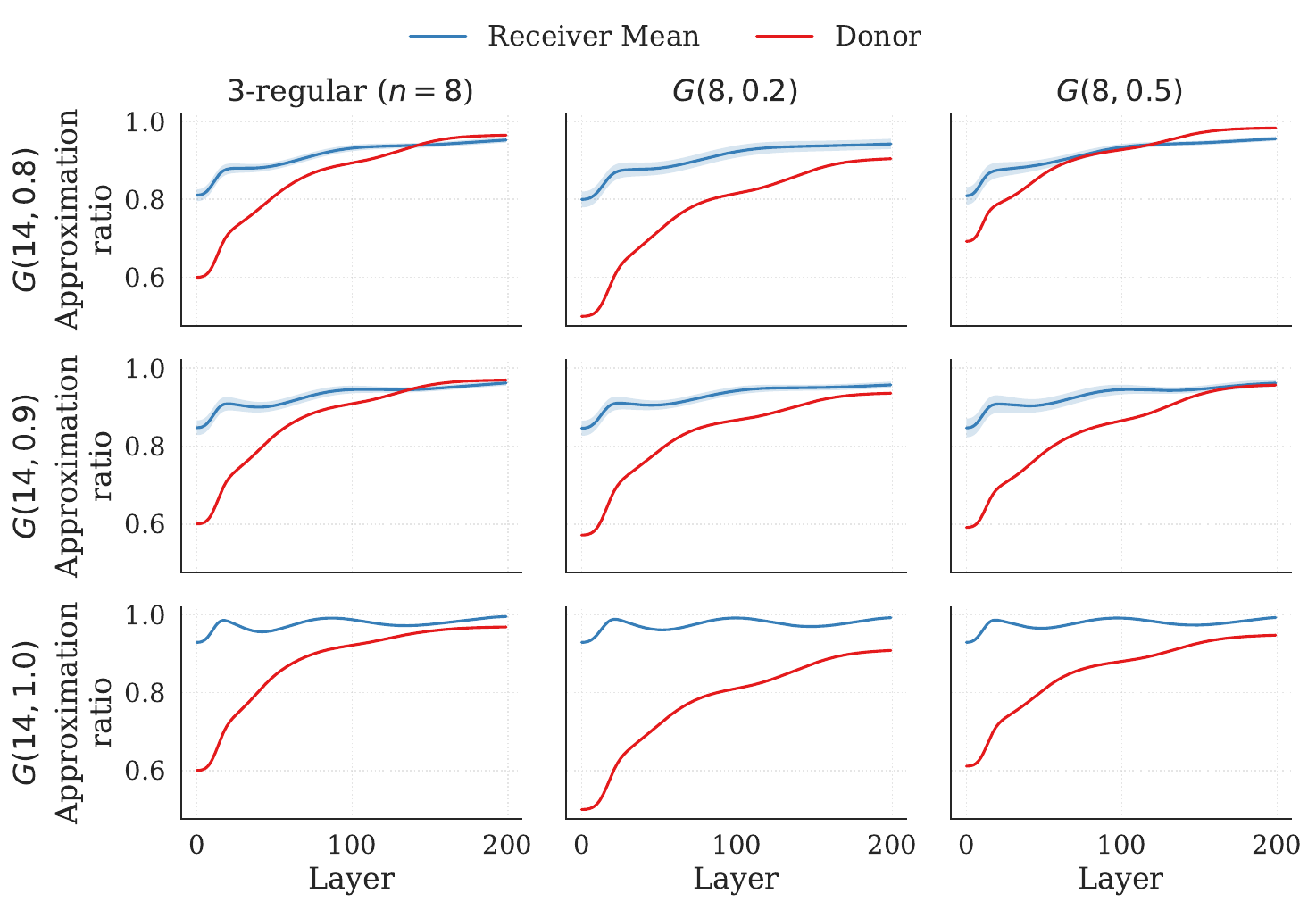}
    \caption{Representative transfer trajectories for dense 14-node Erd\H{o}s--R\'enyi recipients ($p \in \{0.8, 0.9, 1.0\}$) from 8-node donor graphs. The transferred mean approximation-ratio curves (blue) exhibit narrow variance and monotonic growth, consistently matching or exceeding the donor training trajectories (red).}
    \label{fig:dense_er_transfer}
\end{figure}

\paragraph{Transfer is more effective for dense Erd\H{o}s--R\'enyi recipient graphs.}
As shown in Figure~\ref{fig:dense_er_transfer}, transferred parameter sequences achieve high approximation ratios on dense 14-node Erd\H{o}s--R\'enyi recipients, frequently matching or exceeding the donor-side training trajectory. This stability is most pronounced for recipient densities $p \in \{0.8, 0.9, 1.0\}$. In these cases, the final approximation ratio reaches approximately 0.95 for $p=0.8$, 0.96 for $p=0.9$, and 0.98 for the complete graph ($p=1.0$). Crucially, this high performance is achieved with remarkably low variance across the recipient instances, demonstrating that transfer to structurally homogeneous, dense graphs is not only effective but highly reliable.

This behavior can be interpreted by translating the quantum dynamics of the FALQON commutator into the structural properties of the target graphs. In FALQON, the commutator expectation value $\langle i[H_M, H_C] \rangle$ acts as a feedback signal—analogous to a gradient in classical optimization—that dictates the next parameter update. For highly dense graphs, the adjacency matrix is heavily populated, meaning local structural irregularities (such as specific missing edges) have a vanishingly small impact on the overall cost function. The optimization landscape becomes uniform, governed by the average degree of the nodes rather than specific local connectivities. Because of this structural homogeneity, the feedback signals generated by a small dense donor proportionally mirror the macroscopic behavior of a larger dense recipient. Consequently, the control parameters learned on the smaller instance remain optimal for the larger one, bypassing the need for instance-specific recalculation.

\begin{figure}[htb]
    \centering
    \includegraphics[width=\columnwidth]{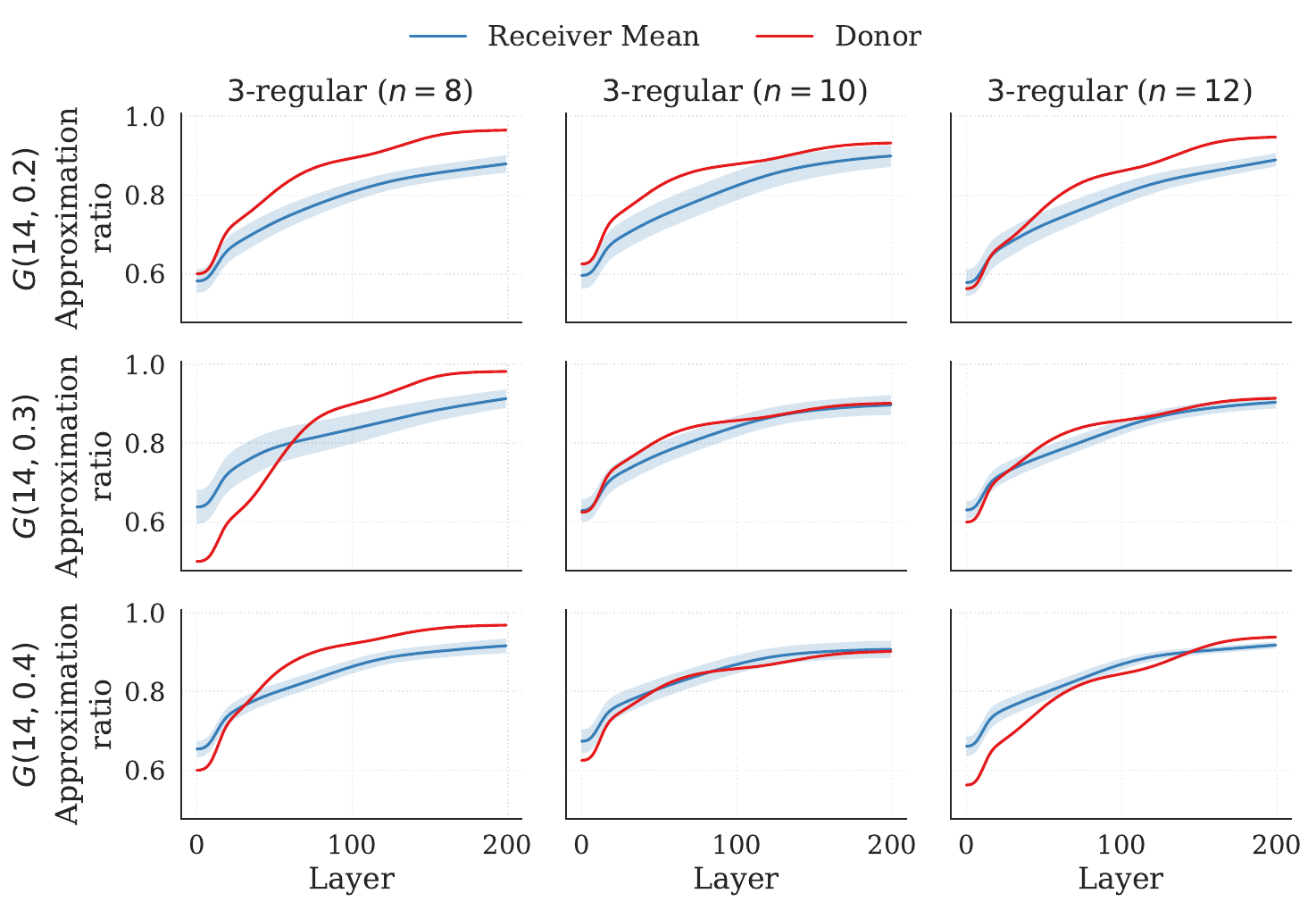}
    \caption{Representative transfer trajectories for sparse 14-node Erd\H{o}s--R\'enyi recipients ($p \in \{0.2, 0.3, 0.4\}$) from 3-regular donor graphs of sizes 8, 10, and 12. The transferred curves (blue) remain below the donor references (red) for $p=0.2$, with the gap narrowing as recipient density increases.}
    \label{fig:sparse_cross_family}
\end{figure}

\paragraph{Sparse cross-family transfer is substantially harder.}
A contrasting pattern appears in cross-family transfers from 3-regular donor graphs to sparse 14-node Erd\H{o}s--R\'enyi recipients (Figure~\ref{fig:sparse_cross_family}). For highly sparse recipients ($p=0.2$), transferred curves consistently remain below the donor-side reference. As recipient density increases to $p=0.3$ and $p=0.4$, this performance gap narrows. This sensitivity indicates that highly sparse target graphs present greater structural variability and a stronger mismatch with 3-regular donors. Consequently, sparse cross-family settings act as a stress test, revealing that transfer quality depends heavily on the structural compatibility between source and target Hamiltonians.

\begin{figure}[htb]
    \centering
    \includegraphics[width=\columnwidth]{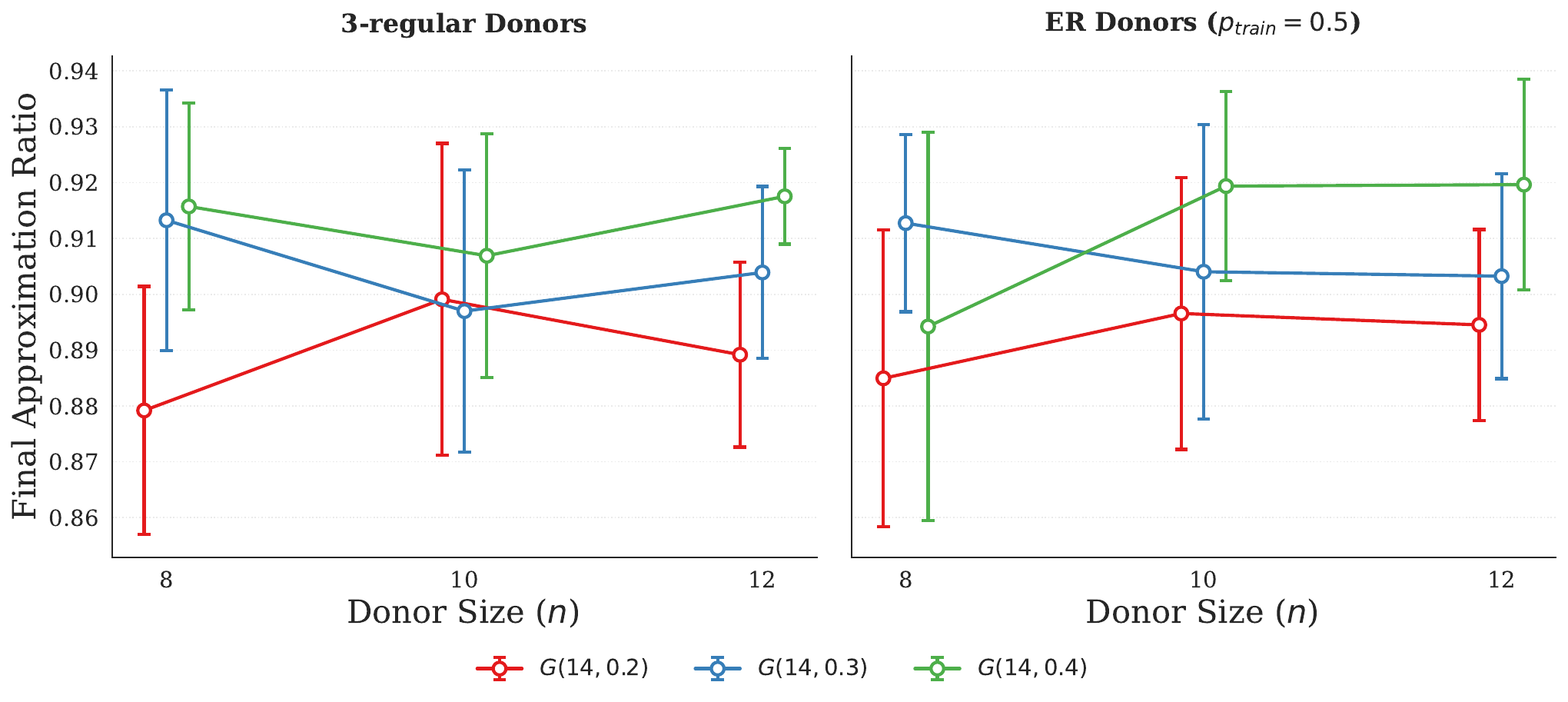}
    \caption{Comparison of transfer performance across donor sizes for (a) 3-regular and (b) Erd\H{o}s--R\'enyi donors (trained at $p=0.5$). The substantial overlap in error bars indicates that transfer quality is resilient to donor size, achieving highly competitive and statistically comparable performance even with 8-node donors.}
    \label{fig:donor_size_effect}
\end{figure}

\paragraph{Transfer performance is resilient to donor size.}
The transfer behavior illustrated in Figure~\ref{fig:donor_size_effect} demonstrates significant resilience across donor sizes. For both 3-regular and Erd\H{o}s--R\'enyi ($p=0.5$) training families, the final approximation ratios for 14-node recipients remain stable across $n \in \{8, 10, 12\}$. While mean performance slightly fluctuates across these donor sizes, the substantial overlap in their standard deviations indicates no statistically significant advantage to using larger donor instances. This confirms that the smallest, most computationally efficient 8-node donors capture control features robust enough for effective reuse, yielding performance statistically indistinguishable from larger donors.

Taken together, these results indicate that parameter transfer in FALQON is primarily governed by the structural properties of the recipient graph rather than the donor size. Transfer quality and reliability improve significantly for denser recipient graphs, while sparse cross-family settings pose the greatest challenge. Crucially, the robust performance of 8-node donors—validated by overlapping variance bounds across scales—suggests that FALQON learns scalable control sequences independent of specific training instance sizes. This supports the viability of leveraging small, easily optimized donors to accelerate optimization on larger target Hamiltonians.

\section{Conclusion}
\label{sec:conclusion}

This work presented an empirical study of parameter transfer in FALQON for Max-Cut on random graphs. We investigated whether parameter sequences learned on smaller donor graphs can be reused on moderately larger 14-node recipient graphs across different graph families and densities. The empirical results reveal three main behaviors. First, transferring to an easier problem instance yields strong results: as the Erd\H{o}s--R\'enyi recipient approaches a complete graph, parameters from virtually any donor achieve high approximation ratios. This is likely because cutting a nearly complete graph is an inherently easier Max-Cut instance. Second, transferring to a harder problem results in degraded performance: sparse cross-family transfer from 3-regular donors to highly sparse Erd\H{o}s--R\'enyi recipients remains substantially challenging.

Third, the success of the transfer is primarily dictated by the recipient graph, with the donor having minimal impact. Transfer performance is remarkably resilient to the size gap; parameter sequences learned on smaller 8-node donor graphs transfer competitively to 14-node recipients when compared to those learned on 10- and 12-node donors. This resilience indicates that the specific scale of the donor instance is a secondary factor. Ultimately, if the target graph is an easy instance, the transfer will succeed regardless of the donor.

Overall, these findings indicate that parameter transfer is a highly promising strategy for reducing the cost of FALQON-based optimization, as computationally cheap, small donor instances can yield effective parameter sequences for larger targets. As future work, it would be valuable to investigate the absolute limits of this scale invariance and whether some form of parameter normalization or rescaling can extend transferability across much more extreme donor–recipient size gaps. Additional directions include extending the analysis to significantly larger graph sizes compared against standard heuristic baselines, evaluating transfer under noisy hardware constraints, and explicitly investigating how the spectral properties and the energy landscape of the problem Hamiltonians dictate transferability bounds, with a particular focus on understanding and overcoming the limitations observed in sparse cross-family regimes.

\section*{Acknowledgements}
\label{sec:acknowledgements}

This work was executed under the TIC26 – Brazil Quantum Camp project, funded within the scope of the Prioritized Informatics Programs and Projects (PPI), Process No. 01245.008254/2025-22, under the responsibility of the Ministry of Science, Technology and Innovation (MCTI), with operational coordination by the Association for the Promotion of Brazilian Software Excellence (SOFTEX), and executed by CESAR and the Instituto de Pesquisas ELDORADO.

\bibliographystyle{sbc}
\bibliography{sbc-template}

\end{document}